\documentclass[prd,twocolumn,nofootinbib,amsmath,amssymb,floatfix,superscriptaddress]{revtex4}
\usepackage{multirow}
\usepackage{graphicx}
\usepackage{epstopdf}
\usepackage{dcolumn}
\usepackage{bm}
\usepackage{threeparttable}
\usepackage{subfigure}
\usepackage{color,txfonts}
\usepackage{ulem}
\definecolor{blue0}{rgb}{0,0,0.6}
\usepackage[colorlinks,linkcolor=blue0,anchorcolor=blue0,citecolor=blue0,urlcolor=blue0]{hyperref}

\newcommand{\beq}{\begin{equation}}
\newcommand{\eeq}{\end{equation}}
\newcommand{\beqa}{\begin{eqnarray}}
\newcommand{\eeqa}{\end{eqnarray}}

\begin{document}

\title{Constraints on Axion-like Particles from the gamma-ray observation of the Galactic Center}
\author{Ben-Yang Zhu}
\affiliation{Key Laboratory of Dark Matter and Space Astronomy, Purple Mountain Observatory, Chinese Academy of Sciences, 210033 Nanjing, Jiangsu, China}
\affiliation{School of Astronomy and Space Science, University of Science and Technology of China, 230026 Hefei, Anhui, China}
\author{Xiaoyuan Huang\footnote{Corresponding author: xyhuang@pmo.ac.cn}}
\affiliation{Key Laboratory of Dark Matter and Space Astronomy, Purple Mountain Observatory, Chinese Academy of Sciences, 210033 Nanjing, Jiangsu, China}
\affiliation{School of Astronomy and Space Science, University of Science and Technology of China, 230026 Hefei, Anhui, China}
\author{Peng-Fei Yin\footnote{Corresponding author: yinpf@ihep.ac.cn}}
\affiliation{Key Laboratory of Particle Astrophysics, Institute of High Energy Physics,
Chinese Academy of Sciences, Beijing 100049, China}

\date{\today}

\begin{abstract}
High energy photons originating from the Galactic Center (GC) region have the potential to undergo significant photon-axion-like particle (ALP) oscillation effects, primarily induced by the presence of intense magnetic fields in this region.
Observations conducted by imaging atmospheric Cherenkov telescopes have detected very high energy gamma-rays originating from a point source known as HESS J1745-290,  situated in close proximity to the GC. This source is conjectured to be associated with the supermassive black hole Sagittarius A$^*$. 
The GC region contains diverse structures, including molecular clouds and non-thermal filaments, which collectively contribute to the intricate magnetic field configurations in this region.
By utilizing a magnetic field model specific in the GC region, we explore the phenomenon of photon-ALP oscillations in the gamma-ray spectrum of HESS J1745-290. Our analysis does not reveal any discernible signature of photon-ALP oscillations, yielding significant constraints that serve as a complement to  gamma-ray observations of extragalactic sources across a broad parameter region. The uncertainties arising from the outer Galactic magnetic field models have minor impacts on our results, except for ALP masses around 10$^{-7}$ eV, as the dominant influence originates from the intense magnetic field strength in the inner GC region.
\end{abstract}

\maketitle

\section{Introduction}
The axion, a light pseudoscalar particle, was initially predicated by the Peccei-Quinn mechanism solving the strong CP problem within quantum chromodynamics (QCD) \cite{qcd1,Peccei:1977hh,qcd3,qcd4}. Many new physics models beyond the standard model also predict the existence of axion-like particles (ALPs), which share similarities with the QCD axion but demonstrate a broader range of couplings and masses \cite{2008LNP...741..115M,1998PhRvL..81.3067C,2010ARNPS..60..405J}.
The QCD axion and ALPs could serve as promising candidates for cold dark matter and may potentially play a significant role in cosmology  ~\cite{1985MNRAS.215..575K,Marsh:2015xka,dmcandidate1,dmcandidate2,dmcandidate3}, thereby prompting extensive investigations in the field of new physics research. 


The interaction between photons and ALPs can lead to photon-ALP oscillations in the presence of an external magnetic field. This phenomenon has been extensively explored through laboratory experiments \cite{Sikivie:1983ip,DePanfilis:1987dk,VanBibber:1987rq,Bradley:2003kg,Rabadan:2005dm,Caldwell:2016dcw,McAllister:2017lkb,CAST:2017uph,ADMX:2019uok,HAYSTAC:2020kwv,QUAX:2020adt,Choi:2020wyr}.
Astrophysical sources also provide an ideal environment for the exploration of this phenomenon, potentially causing irregularities in photon spectra.  High-energy photons interact with lower-energy background photons, leading to the absorption of high-energy photons and consequently a decrease in observed gamma-ray spectra at high energies. However, the conversion of photons into ALPs could potentially alleviate this absorption, offering a promising avenue for ALP detection. Numerous studies have been conducted to impose constraints on the ALP parameters, taking into account these effects
 \cite{Raffelt:1987im, DeAngelis:2007dqd, Hooper:2007bq, Simet:2007sa, Mirizzi:2007hr, Mirizzi:2009aj, Belikov:2010ma, DeAngelis:2011id, Horns:2012kw, HESS:2013udx, Meyer:2013pny, Tavecchio:2014yoa, Meyer:2014epa, Meyer:2014gta, Fermi-LAT:2016nkz, Meyer:2016wrm, Berenji:2016jji, Galanti:2018upl, Galanti:2018myb, Zhang:2018wpc, Liang:2018mqm, Xia:2018xbt, Majumdar:2018sbv, Guo:2020kiq, Li:2020pcn, Li:2021gxs, Cheng:2020bhr, Liang:2020roo, Bi:2020ths,Xia:2019yud, Dessert:2022yqq, Kohri:2017ljt,  Davies:2022wvj,Gao:2023dvn,Gao:2023und,Li:2024ivs, MAGIC:2024arq}. 

Previous studies on astrophysical photon-ALP oscillations at high energies have predominantly focused on extragalactic sources. Photons emitted from these sources can travel long distances through magnetic fields, resulting in observable oscillation phenomena. In contrast, photons emitted from Galactic sources may exhibit limited oscillation effects \cite{Xia:2018xbt,Majumdar:2018sbv,Xia:2019yud,Bi:2020ths,Li:2024ivs}, as they travel shorter distances within the Galactic magnetic field (GMF). However, photons originating from the Galactic Center (GC) region could potentially exhibit substantial oscillation effects that are detectable.

The GC is one of the most complex and active regions within the Milky Way. At its core resides a supermassive black hole known as Sagittarius A$^*$ (Sgr A$^*$), boasting a mass of $\sim4.154\pm0.014\times10^{6}{\ \rm M_\odot}$ \cite{sgrmass}. High-energy gamma rays have been observed in this region, encompassing extended emissions from the central molecular zone (CMZ) and emissions from the central point source, HESS J1745-290, potentially associated with Sgr A$^*$ \cite{HESS2016,HESS:2017tce,veritas,magic,4FGL,Huang:2020ngv,2024arXiv240516348M}. Furthermore, various observations have indicated the presence of a robust magnetic field within the CMZ \cite{2009A&A...505.1183F,2019A&A...630A..74M,2022MNRAS.513.3493H}. This magnetic field may play an important role in the dynamics of the interstellar medium, as well as in processes such as star formation and cosmic-ray propagation~\cite{2010Natur.463...65C,Kruijssen:2013jva,2022arXiv220708097B,Dorner:2023dqd,2024ApJ...963..130B,2024arXiv240313048T,2024ApJ...962...39L}. 
Although many aspects of the magnetic field configuration in the GC region remains uncertain, a magnetic field model in the CMZ has been constructed based on observational data of diffuse gas, non-thermal radio filaments, and molecular clouds \cite{Guenduez2020}. This analytical depiction of the magnetic field holds promise for investigating cosmic-ray and gamma-ray phenomena in the GC region~\cite{BeckerTjus:2020xzg,2022arXiv220708097B,AlvesBatista:2022vem,Dorner:2023dqd}.

In this study, we utilize the HESS observations of the central point source, HESS J1745-290, situated at the GC to search for photon-ALP oscillations within the gamma-ray spectrum and to impose constraints on the ALP parameters. While some studies have established constraints on the ALP parameters through high energy gamma-ray observations of Galactic
sources \cite{Xia:2018xbt,Majumdar:2018sbv,Xia:2019yud,Bi:2020ths,Li:2024ivs}, the unique characteristics of HESS J1745-290 make it an exceptional target for detecting photon-ALP oscillations, particularly owing to the intense magnetic field in the CMZ. Furthermore, for accurate computation of photon-ALP conversion along the path from the GC to Earth, it is essential to consider contributions from the outer regions beyond the CMZ. Recent advancements in measuring Faraday rotation of extragalactic sources and Galactic synchrotron emissions have facilitated  continuous improvements in the GMF model~\cite{J12,pshirkov,TF17,Unger:2023lob}.  Our analysis investigates the impacts of the GMF model and the uncertainty arising from the location of HESS J1745-290 on the ALP constraints. 

The paper is organized as follows. In Section.~\ref{secII}, we introduce the magnetic field configuration in the Milky Way and gamma-ray observations of the central point source at the GC. Section.~\ref{sec3} outlines the method for estimating photon-ALP oscillations, and describes the statistical approach employed to derive constraints on the ALP parameters. The constraints derived from the GC observations, along with discussions on uncertainties and implications, are presented in Section~\ref{sec4}. Finally, we summarize our results in Section~\ref{sec5}.

\section{magnetic fields and gamma-ray observations}
\label{secII}

\subsection{Magnetic fields in the GC and outer regions}

\begin{figure}
    \centering
    \includegraphics[width=0.45\textwidth]{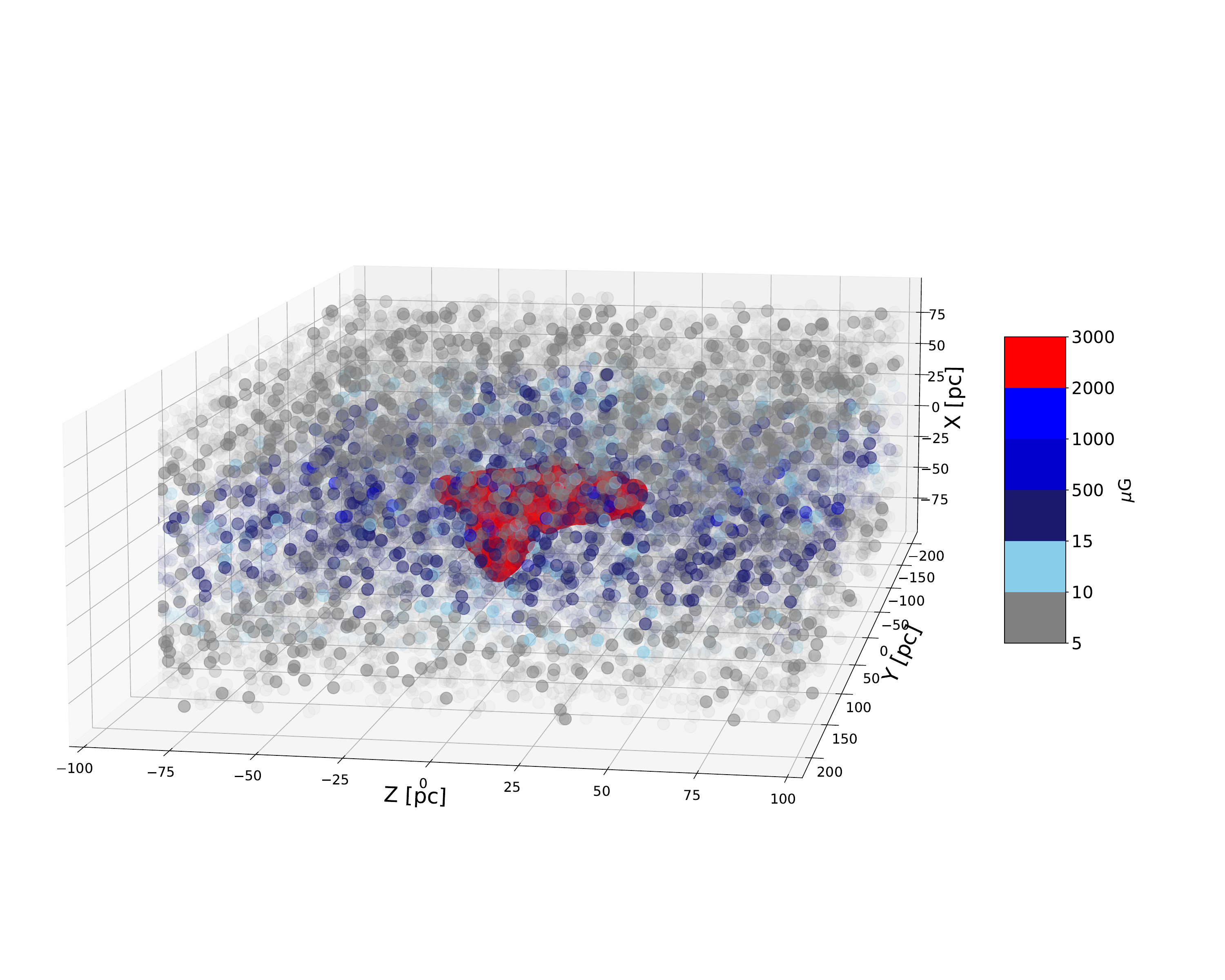}
    \includegraphics[width=0.45\textwidth]{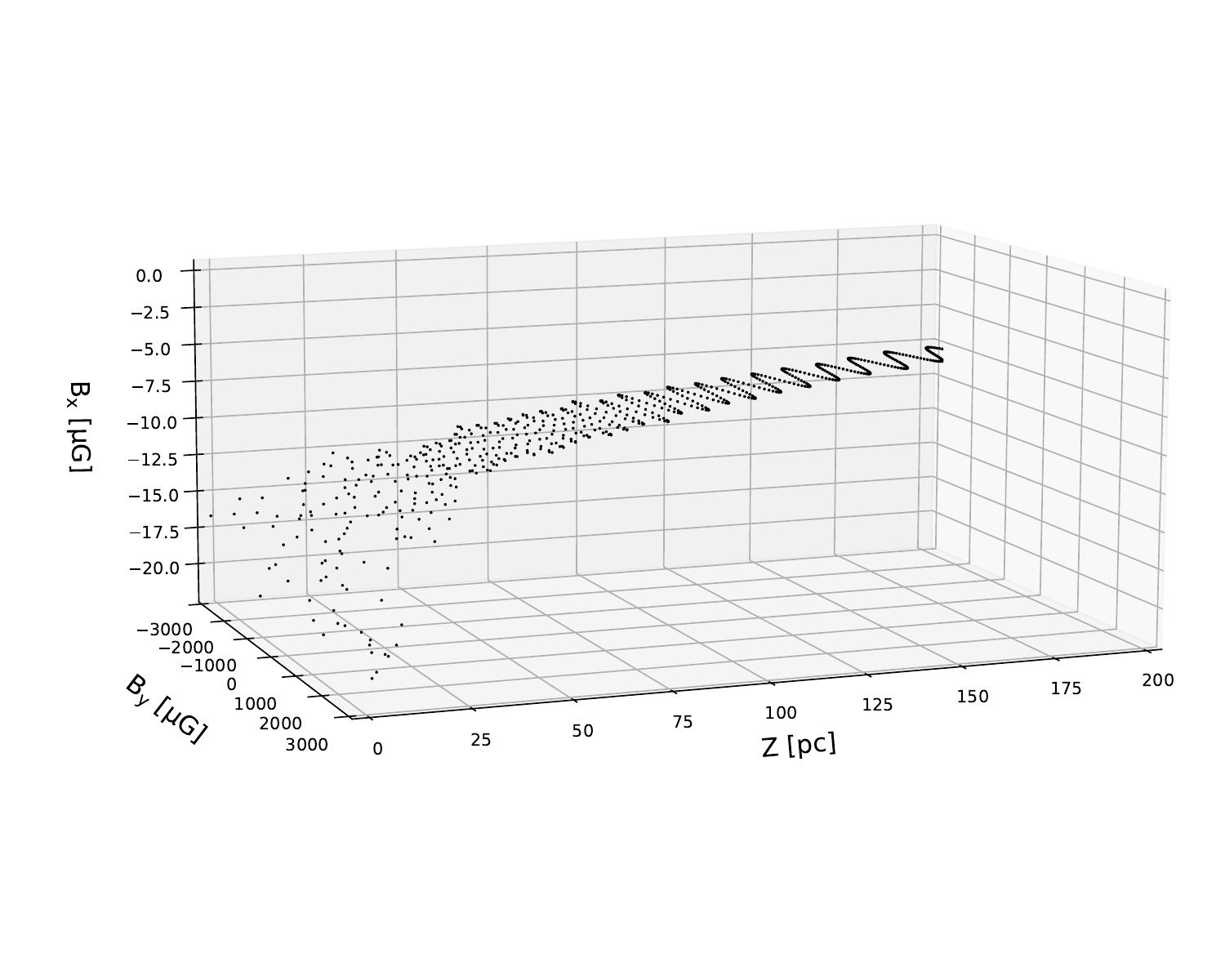}
   \caption{{The magnetic field in the CMZ under consideration. Upper panel:  the distribution of magnetic field strength in the CMZ. The z axis aligns with the line extending from the GC to Earth. Lower panel: the magnetic field components in the transverse plane along the line extending from the central point source to Earth. The z axis aligns with this line. In both panels, the y axis corresponds to the direction of $l=90^\circ$, where $l$ represents the Galactic longitude.}}
    \label{CMZfield}
\end{figure}

GMFs play an important role in various astrophysical processes, including the evolution of molecular clouds, star formation, and cosmic-ray propagation. The configuration of these magnetic fields would also significantly affect the phenomenon of photon-ALP oscillations. Although detailed information about GMFs is somewhat limited, considerable efforts have been dedicated to constrain their parameters and structure through measurements of Faraday rotation of extragalactic sources and Galactic synchrotron emissions \cite{J12,pshirkov,TF17,Unger:2023lob}. Various GMF models have been developed, yet accurately modeling magnetic fields in intricate regions like the Galactic center, particularly the CMZ, poses challenges. Recently, a novel analytical model has been proposed to describe the magnetic field configuration in the central 200 pc of the CMZ~\cite{Guenduez2020}.
By combining observations of diffuse gas, non-thermal radio filaments, and molecular clouds with theoretical magnetic field models, a magnetic field configuration has been derived to capture the main features of polarization maps. Some properties of this magnetic field configuration are shown in Fig.~\ref{CMZfield}. The model indicates that the typical magnetic field strength in the CMZ could be an order of magnitude higher than that in the Galactic disk, and could even reach $\mathcal{O}(10)$ mG in molecular clouds. This model complements existing GMF models, providing a more comprehensive understanding of the magnetic field structure in the CMZ. This detailed knowledge is crucial for exploring the potential impact of photon-ALP oscillations on the gamma-ray spectrum of the central point source HESS J1745-290. 

The strength of photon-ALP oscillations is influenced by the transverse components of all the magnetic fields along the line-of-sight, extending from the source to the observer. Therefore, it is important to consider both the magnetic field configurations in the CMZ and the Galactic disk. In this study, we utilize the magnetic field configuration in the CMZ from Ref.~\cite{Guenduez2020}, and include a superposition of the GMF in the outer region proposed by Ref.~\cite{J12} as the baseline model. The GMF model of Jansson \& Farrar \cite{J12} (hereafter referred to as JF12) is constructed using observations of Faraday rotation from extragalactic sources and Galactic synchrotron emissions. This model encompasses the disk, halo, and X-shaped field components. However, it is noteworthy that this model sets the disk component to zero within Galactic radii of 3 kpc, which may potentially result in a conservative estimation of photon-ALP oscillations along the line-of-sight\footnote{While the survival probability can be roughly estimated as $\propto B^2d^2$, the complex interplay of the changing orientations and amplitudes of the transverse components of the magnetic fields along the line of sight would complicate the effect.}. In addition to this GMF model, some other models have been introduced as well~\cite{pshirkov,TF17, Unger:2023lob}. 
Ref.~\cite{pshirkov} presents a GMF model that incorporates both disk and halo components, which is fitted using rotation measures of extragalactic radio sources. This model includes the ASS and BSS versions, distinguished by the field direction in two distinct arms. Ref.~\cite{TF17} introduces six best-fit GMF models (C0Ad1, C0Bd1, C0Dd1, C1Ad1, C1Bd1, and C1Dd1) corresponding to axisymmetric and bisymmetric halo fields, along with diverse Galactic disk models. By exploring these distinct GMF models, we aim to address the uncertainties associated with GMF modeling and enhance the accuracy of our analyses.

In the study of Ref.~\cite{Guenduez2020}, the magnetic field configurations are independently fitted in the CMZ without other regions of the Galaxy.  This approach results in a discontinuity in the magnetic field modeling at the boundaries where the CMZ interfaces with the GMF. Specifically, the background field in the intercloud medium (ICM) within the CMZ does not maintain continuous differentiability or the divergence-free property at these interfaces.
While it is ideal to adjust the magnetic field in the ICM within the CMZ while ensuring an average value of approximately 10 $\mu$G, the absence of magnetic field models that comprehensively cover the entire Milky Way poses a challenge beyond the scope of our current investigation. In our analyses, we have considered scenarios with and without the ICM in the CMZ region. The results indicate that this choice has a negligible impact, given the relatively small size of the CMZ and the relatively low magnetic field strength of the ICM. In order to obtain a complete magnetic field configuration, we simply superpose the magnetic field in the CMZ with the GMF. For the magnetic field in the CMZ and the GMF model from Ref.~\cite{TF17}, we utilize the settings of the CRPropa code  \footnote{\url{https://github.com/CRPropa/CRPropa3/blob/3.2.1/src/magneticField/CMZField.cpp}} \cite{AlvesBatista:2022vem} to describe their configurations. In addition, we employ the settings of the gammaALPs package\footnote{\url{https://gammaalps.readthedocs.io/en/latest/\#}} \cite{Meyer:2021pbp} to describe other GMF models.

\subsection{Gamma-ray observations of the central point source}
The central point source at the GC has been detected by various observatories, including the Fermi Large Area Telescope (Fermi-LAT) \cite{4FGL}, H.E.S.S. \cite{HESS2016}, MAGIC \cite{magic}, and VERITAS \cite{veritas} in the high-energy and very high-energy (VHE) ranges. This source may be associated with the supermassive black hole Sgr A$^*$. Observations from Fermi-LAT indicate that the spectrum of this source can be characterized by a broken power law, with spectral indices of $2.00\pm0.04$ up to 3 GeV and $2.68\pm0.05$ beyond this energy~\cite{4FGL}. Additionally, the high-quality observation of the source HESS J1745-290 by H.E.S.S. presents a deviation from a simple power-law fitt. Instead, the H.E.S.S. data suggest an  exponential cutoff power-law spectrum with a cutoff around 10 TeV with an index of $\sim 2.14$. 

The presence of a distinct spectral break in the high energy and VHE observations has prompted extensive investigations into the gamma-ray emission mechanisms of the point source at the GC \cite{GCdiff,Malyshev:2015hqa,hybrid}. In addition to this source, another significant gamma-ray component in the GC vicinity is the so called Galactic Center excess, which is predominantly observed in the GeV energy range \cite{gce1,Fermi-LAT:2015sau,Fermi-LAT:2017opo}. The high-energy tail of this component can have a substantial impact on the spectral shape of sources located in the GC region, from ten GeV to even hundreds of GeV \cite{Huang:2020ngv}. This suggests that the actual spectrum of the central point source in the GeV band may differ from previous analyses. Furthermore, it is important to note that observations from imaging atmospheric Cherenkov telescopes have provided consistent results regarding the spectral index and cutoff energy when assuming an exponential cutoff power-law spectrum. However, significant variations in flux normalization may be observed among these experiments~\cite{veritas}. Notably, H.E.S.S., with its broader energy coverage for the central source, is well-suited for observations of the GC compared to MAGIC and VERITAS. Consequently, this study exclusively focuses on the observational results of the point source HESS J1745-290 from H.E.S.S., while disregarding the results obtained from other experiments.

\section{Analysis}
\label{sec3}
\subsection{Photon-ALP oscillations} 
The interaction between the ALP and photons can be described by the Lagrangian 
\begin{equation}
    \mathcal{L}_{a\gamma}=-\frac{1}{4}g_{a\gamma}F_{\mu\nu}\widetilde{F}^{\mu\nu}a=g_{a\gamma}a\vec{E}\cdot\vec{B},
\end{equation}
where $g_{a\gamma}$ is the coupling coefficient, $a$ represents the ALP field, $F_{\mu\nu}$ and $\widetilde{F}_{\mu\nu}$ are the electromagnetic tensor and its dual respectively, $\vec{E}$  is the photon electric field, and $\vec{B}$ is the magnetic field. The equation of motion
\begin{equation}
\left(i\frac{d}{dz} + E + \mathcal{M}\right) \Psi(z)=0,
\label{Eq:propagation}
\end{equation}
describes the propagation of a monochromatic photon-ALP beam along the $z$ direction in a homogeneous magnetic field \cite{1988PhRvD..37.1237R}, where $z$ is the distance along the propagation direction, and $E$ is the energy of the beam. The status of the beam is described by $\Psi\equiv(A_{1}, A_{2}, a)^T$, where $A_{1}$ and $A_{2}$ represent two polarization amplitudes of the photon. 

The matrix $\mathcal{M}$ describes the ALP oscillations and the absorption of high-energy photons. 
Assuming the transverse magnetic field $B_t$ parallel to $A_2$, $\mathcal{M}$ can be written as \cite{2008LNP...741..115M} 
      \begin{equation}
\mathcal{M}^{(0)} =   \begin{pmatrix}
    \Delta_{\perp}&0&0\\
    0&\Delta_{\parallel}&\Delta_{a\gamma}\\
    0&\Delta_{a\gamma}&\Delta_a\\
      \end{pmatrix},
\end{equation}  where the elements $\Delta_{\perp}$ and $\Delta_{\parallel}$ in matrix arise from photons propagating in a plasma and the QED vacuum polarisation effect, respectively. They can be expressed as:
\begin{equation}
 \label{eq4_1}
    \Delta_{\parallel} = \Delta_{\rm pl} + \frac{7}{2}\Delta_{\rm QED} + \Delta_{\gamma\gamma}-\frac{1}{2}{\rm i}\Gamma_{\rm BG},
  \end{equation}
\begin{equation}
 \label{eq4_2}
    \Delta_{\perp} = \Delta_{\rm pl} + 2\Delta_{\rm QED} + \Delta_{ \gamma\gamma}-\frac{1}{2}{\rm i}\Gamma_{\rm BG}.
  \end{equation}
For an electron density $n_{\rm el}$, the plasma contribution is $\Delta_{\rm pl} = -\omega_{\rm pl}/2E$ with the plasma frequency of $\omega_{\rm pl}^2=4\pi e^2n_{\rm el}/m_e$. The QED term reads $\Delta_{\rm QED} = \alpha E    (B/B_{\rm cr})^2/45\pi$, where $\alpha$ is the fine-structure constant and $B_{\rm cr}$ is the critical magnetic field given by $B_{\rm cr}\equiv m_e^2/e\simeq 4.4\times 10^{13}$ G. $\Delta_{\gamma\gamma}=44\alpha^2E\rho_{\rm RF}/(135m_e^4)$ represents the dispersion effect from photon-photon scattering on environmental radiation field \cite{absorption}. The other two elements, $\Delta_a$ and $\Delta_{a\gamma}$, describing the mass and mixing effects of ALPs respectively, are given by:
\begin{equation}
       \Delta_a = -\frac{m^2_a}{2E},
\end{equation}
\begin{equation}
       \Delta_{a\gamma} = \frac{g_{a\gamma}B_t}{2},
\end{equation}
where $m_a$ is the ALP mass.

Considering the energy range observed by imaging atmospheric Cherenkov telescopes, where very high-energy photons interact with background photon fields, such as the interstellar radiation field and cosmic microwave background, through the $\gamma\gamma$ pair production process, the absorption effect may have considerable effects \cite{2006ApJ...640L.155M,2015JCAP...10..014E,2018PhRvD..98d1302P}. The background absorption term $\Gamma_{\rm BG}$ in Eq.~(\ref{eq4_1}) and Eq.~(\ref{eq4_2}) can be calculated as 
\begin{equation}
    \Gamma_{\rm BG} = \frac{1}{2}\int dE_{\rm BG}\frac{dn_{\rm BG}}{{dE_{\rm BG}}}\hat{\sigma},
\end{equation}
where $E_{\rm BG}$ and $n_{\rm BG}$ represent the energy and number density of the background photons, respectively. The term $\hat{\sigma}$ can be expressed as 
\begin{equation}
    \hat{\sigma}=\int_0^2dx\frac{x}{2}\sigma_{\gamma\gamma},
\end{equation}
where $x=1-\cos\theta$, $s = 2E_\gamma E_{\rm BG}x$ is the square of the center-of-mass energy, $m_e$ is the electron mass, and $\theta$ is the angle between the directions of the interacting photons. The velocity of the $e^\pm$ in the center-of-mass frame is $\beta=\sqrt{1-4m_e^2/s}$, and the cross-section for pair production is given by \cite{2016PhRvD..94f3009V}
\begin{equation}
    \sigma_{\gamma\gamma}=\frac{3}{16}\sigma_T(1-\beta^2)\left[(3-\beta^4)\ln\frac{1+\beta}{1-\beta}-2\beta(2-\beta^2)\right],
\end{equation}
 where $\sigma_T$ represents the Thomson cross-section. The interstellar radiation field model used in this work is adopted from Ref.~\cite{2016PhRvD..94f3009V}. For the GC region, this effect only results in a $\sim$ 1.2\% attenuation of flux at 10 TeV, with even smaller reductions at lower energies.  

While the external magnetic field $B$ is homogeneous, $M$ can be transformed into a diagonal form $D= \mathrm{diag}(D_1, D_2, D_3)$ by employing the similarity transformation $D = W\mathcal{M}^{(0)} W^\dagger$~\cite{2010JCAP...05..010B}, where the rotation matrix is given by
 \begin{equation}
     W = \begin{pmatrix}
         1&0&0\\
         0&\cos\theta&\sin\theta\\
         0&-\sin\theta&\cos\theta\\
     \end{pmatrix},
 \end{equation}
%
with the mixing angle 
\begin{equation}
    \theta=\frac{1}{2}\arctan\left(\frac{2\Delta_{a\gamma}}{\Delta_{\parallel}-\Delta_a}\right).
\end{equation}
The diagonal terms can be explicitly given as: 
\begin{equation}
\label{eq14}
\begin{split}
    &D_1 = \Delta_{\perp},\\
    &D_2 = \frac{(\Delta_\parallel+ \Delta_a)}{2}+\frac{1}{2}\left[(\Delta_a-\Delta_{\parallel})^2+4\Delta_{a\gamma}^2\right]^{1/2},\\
    &D_3 = \frac{(\Delta_{\parallel}+\Delta_a)}{2}-\frac{1}{2}\left[(\Delta_a-\Delta_\parallel)^2+4\Delta^2_{a\gamma}\right]^{1/2}.
\end{split}
\end{equation}

The path of photons traveling from the GC to Earth can be divided into numerous segments, with the magnetic field considered as a constant within each segment. 
Let $\psi$ denotes the angle between $B_{t}$ and $A_2$ in a specific segment.
The mixing matrix can be computed for any orientation of $B_t$, using the similarity transformation 
\begin{equation}
    \mathcal{M} = V^\dagger(\psi)\mathcal{M}^{(0)}V(\psi),
\end{equation}
where the rotation matrix in the transverse plane is given by
\begin{equation}
    V(\psi)= \begin{pmatrix}
        \cos\psi&-\sin\psi&0\\
        \sin\psi&\cos\psi&0\\
        0&0&1
    \end{pmatrix}.
\end{equation}

The generalized density matrix $\rho \equiv \Psi \otimes \Psi^\dagger$ can be used to describe the polarized states of the photon-ALP beam. $\rho$ follows a Von Neumann-like equation \cite{2007PhRvD..76l1301D,2009JCAP...12..004M}:
\begin{equation}
\label{eq21}
    i\frac{d\rho}{dz}=[\rho,\mathcal{M}].
\end{equation} 
The solution to Eq.~(\ref{eq21}) in the $k$-th segment with a size of $\Delta z_{k}$  can be represented as $\rho_{k} = T_{k}\rho_{k-1}T^\dagger_{k}$, where the transfer matrix is calculated as
\begin{equation}
    T_k=V^\dagger(\psi_k)W^\dagger e^{iD
    \Delta z_{k}}WV(\psi_k).
\end{equation}
The transfer matrix $T_k$ can be reformulated as 
\begin{equation}
    T_k = e^{iD_1\Delta z_{k}}T_A(\psi_k)+e^{iD_2\Delta z_{k}}T_B(\psi_k)+e^{iD_3\Delta z_{k}}T_C(\psi_k),
\end{equation}
with the forms of $T_{A,B,C}$ detailed in Ref.~\cite{2010JCAP...05..010B}. The total transfer matrix for all $N$ segments is given by 
\begin{equation}
    T = \prod_{k=1}^N T_k.
\end{equation}
Consequently, the density matrix after propagation can be expressed as 
$\rho_{N} = T\rho_{0}T^\dagger$.

The survival probability of photons across all segments could be calculated as
\begin{equation}
    P_{\gamma\gamma} = {\rm Tr}((\rho_{11}+\rho_{22})T\rho(0)T^\dagger),
\end{equation}
where $\rho_{0}$ = 1/2 diag(1,1,0) is the initial density matrix for initially unpolarized photons, $\rho_{11}$ = diag(1,0,0), and $\rho_{22}$ = diag(0,1,0). Considering the photon-ALP oscillations and high energy photon absorption resulting from background photon fields, the observed spectrum of the source is given by
\begin{equation}
\label{eq22}
    \frac{d\widetilde{N}}{dE} = P_{\gamma\gamma}\frac{dN}{dE}\bigg|_{\rm int},
\end{equation}
where $\frac{dN}{dE}\big|_{\rm int}$ represents the intrinsic spectrum of the source. In this work, the survival probability of photons is calculated by the gammaALPs package.

\subsection{Statistical method}
In order to examine photon-ALP oscillations in the gamma-ray spectrum, we construct the $\chi^2$ function as 
\begin{equation}
\chi^2 = \sum_{\rm i=1}^{\rm N} \frac{\left(\frac{dN}{dE}\bigg|_{i}-\frac{d\widetilde{N}}{dE}\bigg|_{i}\right)^2}{\sigma_i^2},
\end{equation}
where $\frac{d\widetilde{N}}{dE}\bigg|_{i}$, $\frac{dN}{dE}\bigg|_{i}$ and $\sigma_i$ represent the predicted value, observed value, and experimental uncertainty of the photon flux in the $i$-th energy bin, respectively. The intrinsic spectrum is modeled using a power-law with an exponential cutoff, given by $N_0(E/E_0)^{-\Gamma}\exp(-E/E_{\rm cut})$ with the reference $E_0$ fixed at 1 TeV, as outlined in Ref.~\cite{HESS2016}. In the presence of photon-ALP oscillations, the predicted flux should follow the form given by Eq.~(\ref{eq22}) for a specific parameter combination of $(m_a, g_{a\gamma})$. To account for potential oscillation effects that may not be evident in observations due to the limited energy resolution of the detectors, we average the predicted flux within the energy bin.

The test statistic (TS) can be defined as the difference in $\chi^2$ values with and without the photon-ALP oscillation effect, expressed as
\begin{equation}
\label{TS}
    {\rm TS}(m_a,g_{a\gamma}) = \chi^2_{\rm ALP}(\hat{\hat{N_0}},\hat{\hat{\Gamma}},\hat{\hat{E}}_{\rm cut};m_a,g_{a\gamma}) - \chi^2_{\rm Null}(\hat{N_0},\hat{\Gamma},\hat{E}_{\rm cut}),
\end{equation}
where the parameter sets ($\hat{\hat{N_0}}$, $\hat{\hat{\Gamma}}$, $\hat{\hat{E}}_{\rm cut}$) and ($\hat{N_0}$, $\hat{\Gamma}$,$\hat{E}_{\rm cut}$) represent the best-fit parameters for the intrinsic spectrum under the ALP and null hypotheses, respectively. In the case of sufficiently large statistics, the TS statistic is expected to follow a $\chi^2$ distribution~\cite{CL95}. However, due to the non-linear effects of ALPs in the photon spectrum, Wilks' theorem is not applicable here, as discussed in \cite{Fermi-LAT:2016nkz}. Consequently, the TS distribution cannot be adequately described by a $\chi^2$ distribution, necessitating Monte Carlo simulations to derive a more realistic TS distribution. This methodology along with the CLs method \cite{Junk:1999kv,Read:2002hq_cls} widely employed in high-energy experiments, provides a robust approach to establish constraints on the ALP parameters. A more detailed description of this analysis methodology can be found in Ref.~\cite{Gao:2023dvn, Gao:2023und}.

\section{Results and discussions}  

\label{sec4}

\begin{figure}
\centering
    \includegraphics[width=0.45\textwidth]{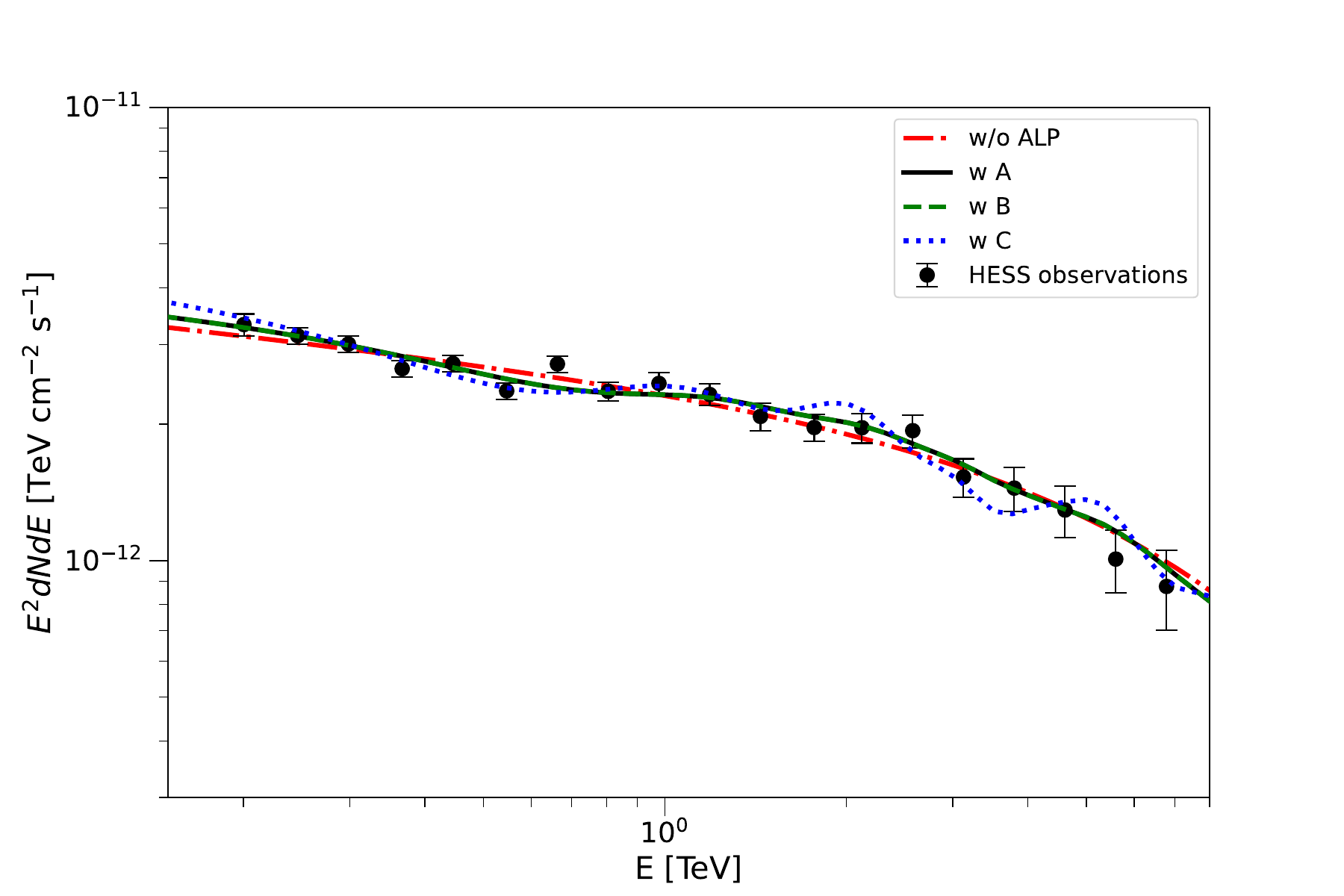}
    \caption{The high energy spectra of HESS J1745-290 under the null hypothesis (solid line) and the ALP
    (dashed line) hypothesis. The three spectra under the ALP hypothesis A, B, and C correspond to $(m_a, g_{a\gamma})= ({10^{-8}\rm
    eV, \;  8\times10^{-11}\rm GeV^{-1}})$,  $(3.3 \times 10^{-8}\rm eV, \;  10^{-11}\rm GeV^{-1})$, and $({3.3\times 10^{-8}\rm eV, \;  3.4\times10^{-11}\rm GeV^{-1}})$, respectively. The black points with uncertainties represent the observational data from H.E.S.S \cite{HESS2016}.}
    \label{sed}
\end{figure}

\begin{figure}
    \centering
    \includegraphics[width=0.45\textwidth]{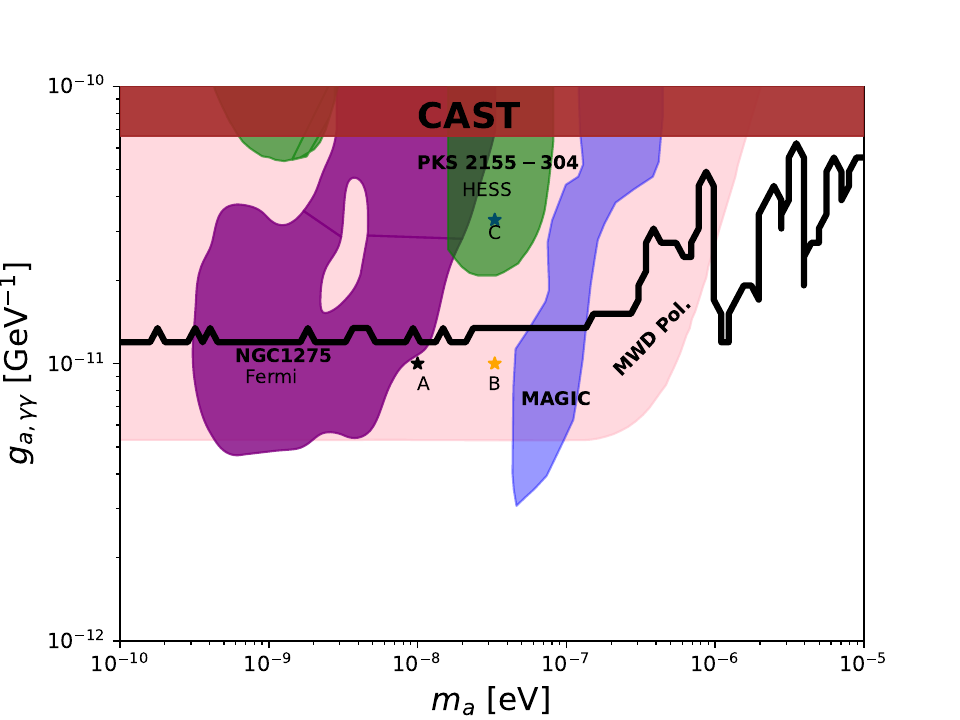}
    \caption{The constraints on the ALP parameters at the 95\% C.L. derived from the observation of HESS J1475-290. The parameter points corresponding to three spectra under the ALP hypothesis shown in Fig.~\ref{sed}, are represented by star symbols. For comparison, the constraints from the CAST experiment \cite{CAST:2017uph}, the H.E.S.S. observation of PKS 2155-304 \cite{HESS:2013udx}, the Fermi-LAT observation of NGC 1275 \cite{Fermi-LAT:2016nkz}, the MAGIC observation of the Perseus cluster \cite{MAGIC:2024arq},  { and the polarization measurements of magnetic white dwarfs \cite{Dessert:2022yqq} are also shown.}}
 \label{c1bd1}
\end{figure}

\begin{figure*}
    \centering
     \includegraphics[width=0.45\textwidth]{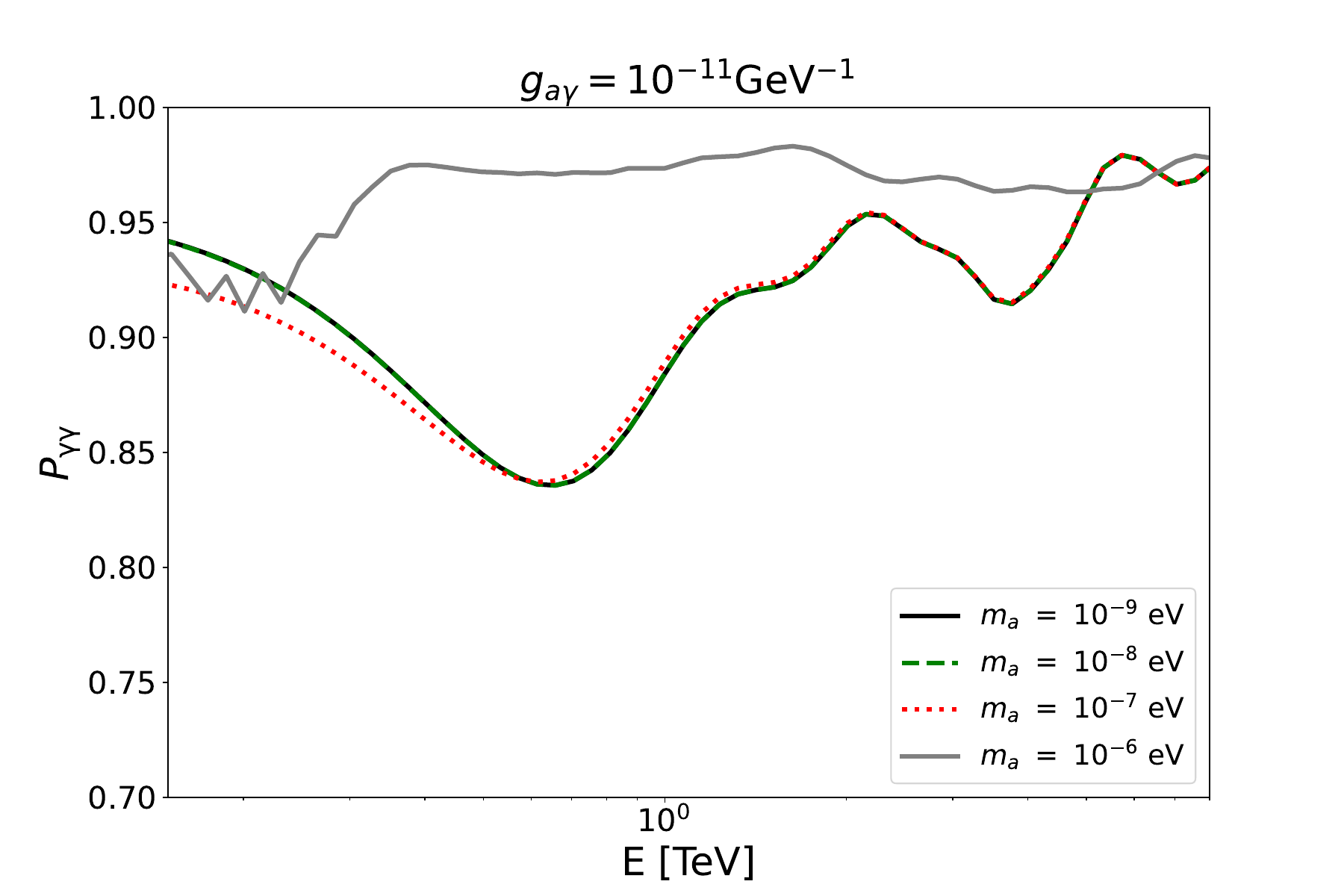} \includegraphics[width=0.45\textwidth]{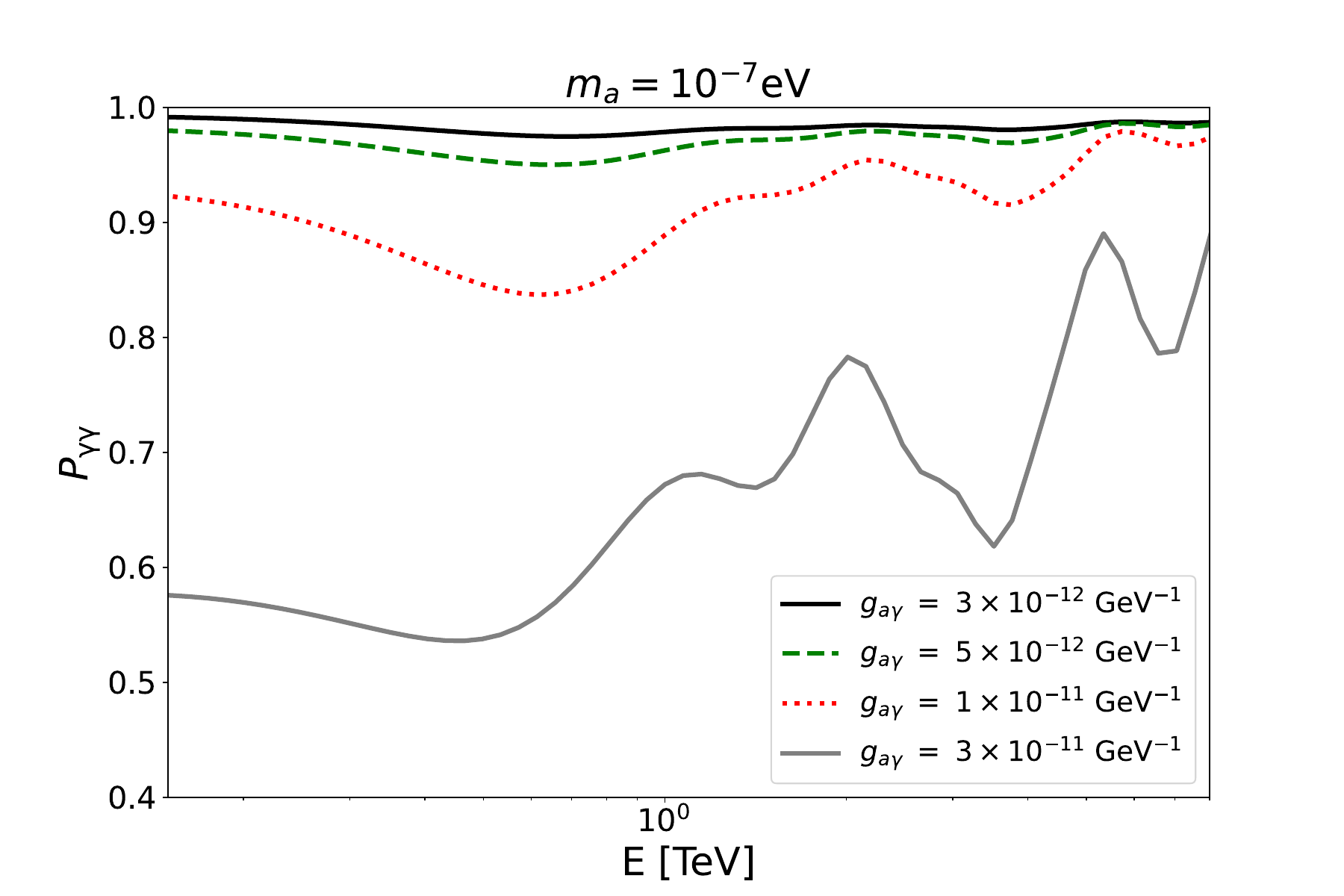}\\
     \includegraphics[width=0.45\textwidth]{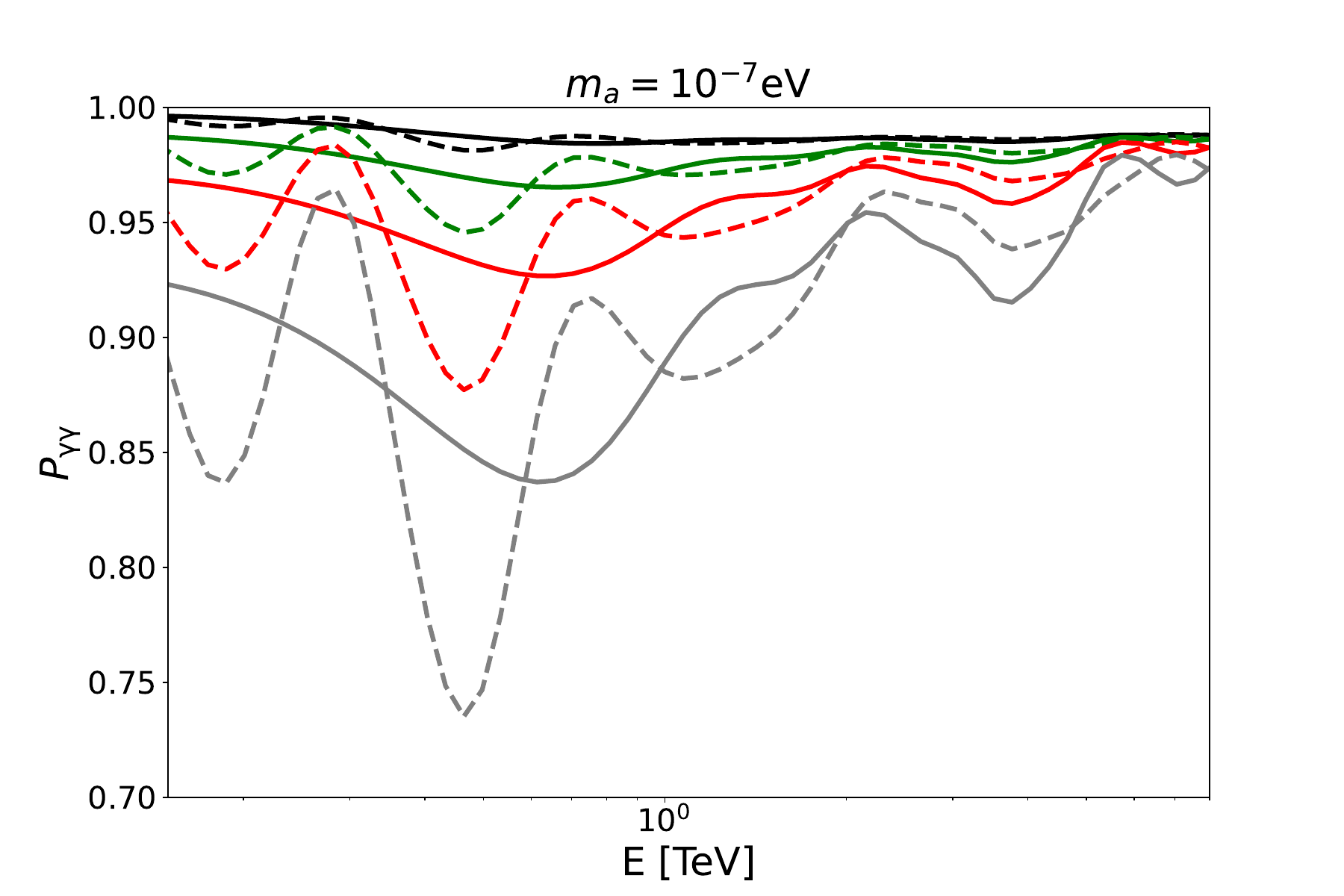}
     \includegraphics[width=0.45\textwidth]{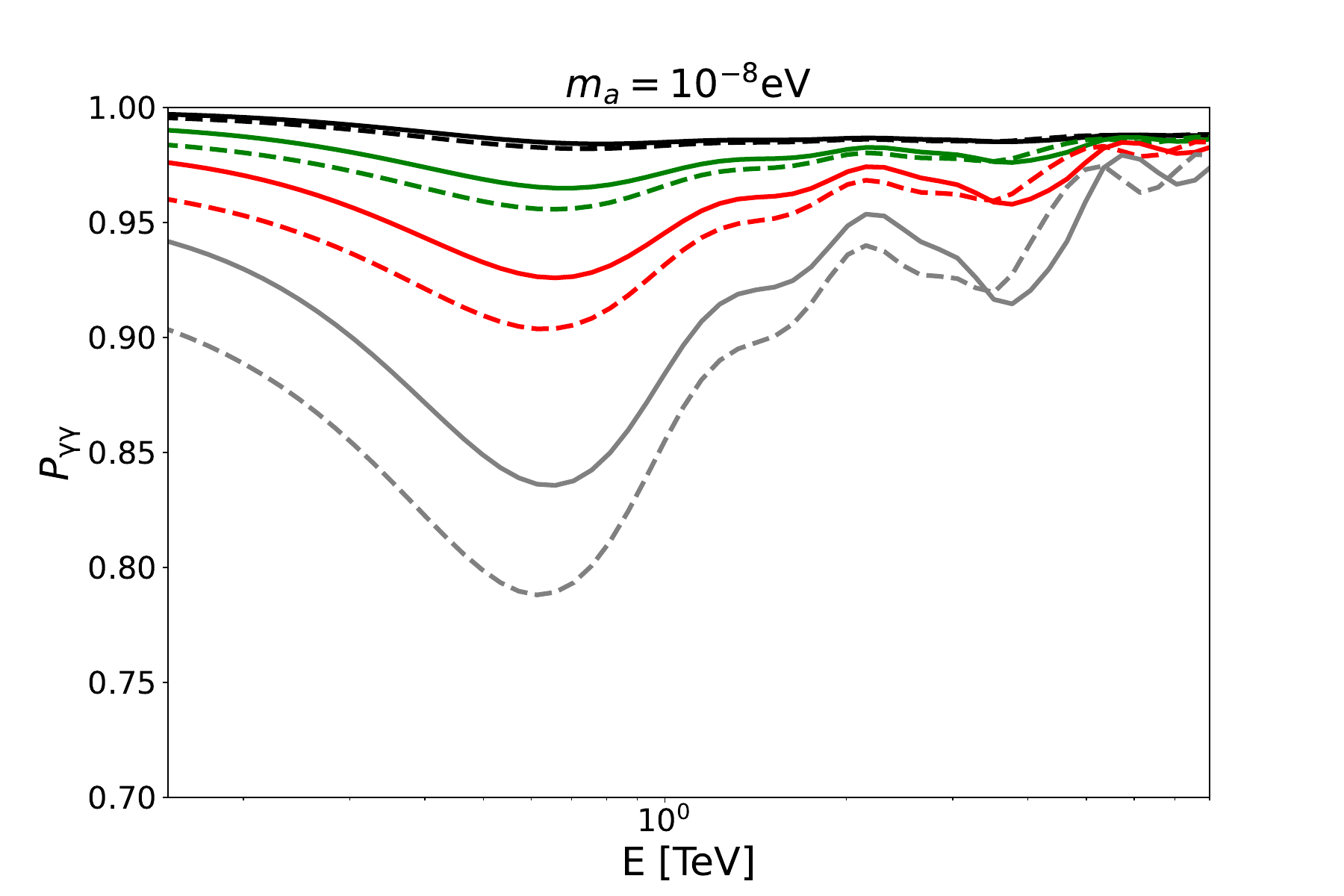}
    \caption{The survival probabilities of photons for different parameter configurations. Upper left panel: results with $g_{a\gamma}=10^{-11}{\rm eV}$ for varying values of $m_a$. Upper right panel: results with $m_a=10^{-7}{\rm eV}$ for varying values of  $g_{a\gamma}$.  Lower left panel: results with $m_a=10^{-7}{\rm eV}$ for varying values of  $g_{a\gamma}$. The solid and dashed lines correspond to results derived with the GMF models of JF12 and C0Bd1, respectively. The black, green, red, and grey lines represent $g_{a\gamma}= 2\times10^{-12},4\times10^{-12},7\times10^{-12}$, and $10^{-11} {\rm\ GeV^{-1}}$, respectively. Lower right panel: similar to the lower left panel, but for $m_a=10^{-8}{\rm eV}$.}
 \label{ratio}
\end{figure*}

In our analysis, we initially calculate the best-fit $\chi^2$ value for the VHE spectrum of HESS J1745-290 under the null hypothesis. This calculation yields a $\chi^2$/d.o.f = 0.84, where d.o.f represents the degrees of freedom. This result indicates that the observed data is well described by the intrinsic spectrum without the presence of ALP effects, as depicted by the red line in Fig.~\ref{sed}. Subsequently, we  investigate the impact of photon-ALP oscillations. We present three spectra with photon-ALP oscillations in Fig.~\ref{sed} to demonstrate how these oscillations can lead to deviations from the intrinsic spectrum for specific parameters. However, upon comparing the fitting results with the null hypothesis, we do not observe a significant improvement in the fit. This suggests that, based on the current data and analysis, there is no strong evidence to support the presence of photon-ALP oscillations in the gamma-ray spectrum of HESS J1745-290.

We perform a systematic scan in the ALP parameter space with $m_a\in[10^{-10},10^{-5}]$ eV and $g_{a\gamma}\in[10^{-12},10^{-10}]\ {\rm GeV^{-1}}$. Utilizing the method outlined in Ref.~\cite{Gao:2023dvn, Gao:2023und}, we establish constraints at a 95\% C.L. and illustrate them in Fig.~\ref{c1bd1}. These constraints are derived based on the CMZ magnetic field model proposed by Ref.~\cite{Guenduez2020} and the JF12 GMF model provided by Ref.~\cite{J12}. For comparison, constraints from other  { astrophysical  observations} are also shown in Fig.~\ref{c1bd1}. Our results are more stringent than the CAST constraints  \cite{CAST:2017uph} across a wide range of masses, { yet they are less stringent than those derived from the observations of magnetic white dwarf polarization \cite{Dessert:2022yqq} for ALP masses below $\sim10^{-6}~$eV. For a more comprehensive overview of constraints on the photon-ALP coupling, interested readers are directed to AxionLimits \footnote{\url{https://cajohare.github.io/AxionLimits/}}.} Notably, our results indicate that the constraints derived from HESS J1745-290, a Galactic source, can complement those obtained from other gamma-ray observations of extragalactic sources  \cite{HESS:2013udx,Fermi-LAT:2016nkz,MAGIC:2024arq} across a broad range of parameters.

The constraints manifest in a nearly mass-independent manner within the low mass region $m_a<10^{-7}$ eV. According to  Eq.~(\ref{eq14}), the variation in the ALP mass term $\Delta_a$ would have a significant impact on the  spectrum unless $\Delta_a \ll \Delta_{\parallel}$. In the context of the strong magnetic field present in the innermost region of the GC region, the term $\Delta_{\parallel}$ is predominantly contributed by the QED term, surpassing the contribution of the mass term in the low mass region. Therefore, mass variations in this region do not affect the constraints significantly. In the upper left panel of Fig.~\ref{ratio}, we show the survival probability of photons $P_{\gamma\gamma}$ with a fixed coupling $g_{a\gamma}= 10^{-11}{\ \rm GeV^{-1}}$ for various ALP masses. 
It is evident that in the low mass range, $P_{\gamma\gamma}$ remains relatively unchanged despite variations in mass. In the upper right panel of Fig.\ref{ratio}, alterations in $P_{\gamma\gamma}$ across different couplings $g_{a\gamma}$ with $m_a=10^{-7}$ eV are presented. A low value of $g_{a\gamma}$ does not significantly modify the mixing matrix, indicating a diminished oscillation effect and rendering these regions in parameter space non-excluded. Conversely, in regions with high coupling values, the significant mixing induces  noticeable oscillations in the spectrum, leading to substantial deviations from the observational data and consequent exclusion of such coupling values.


\begin{figure}
    \centering
    \includegraphics[width=0.45\textwidth]{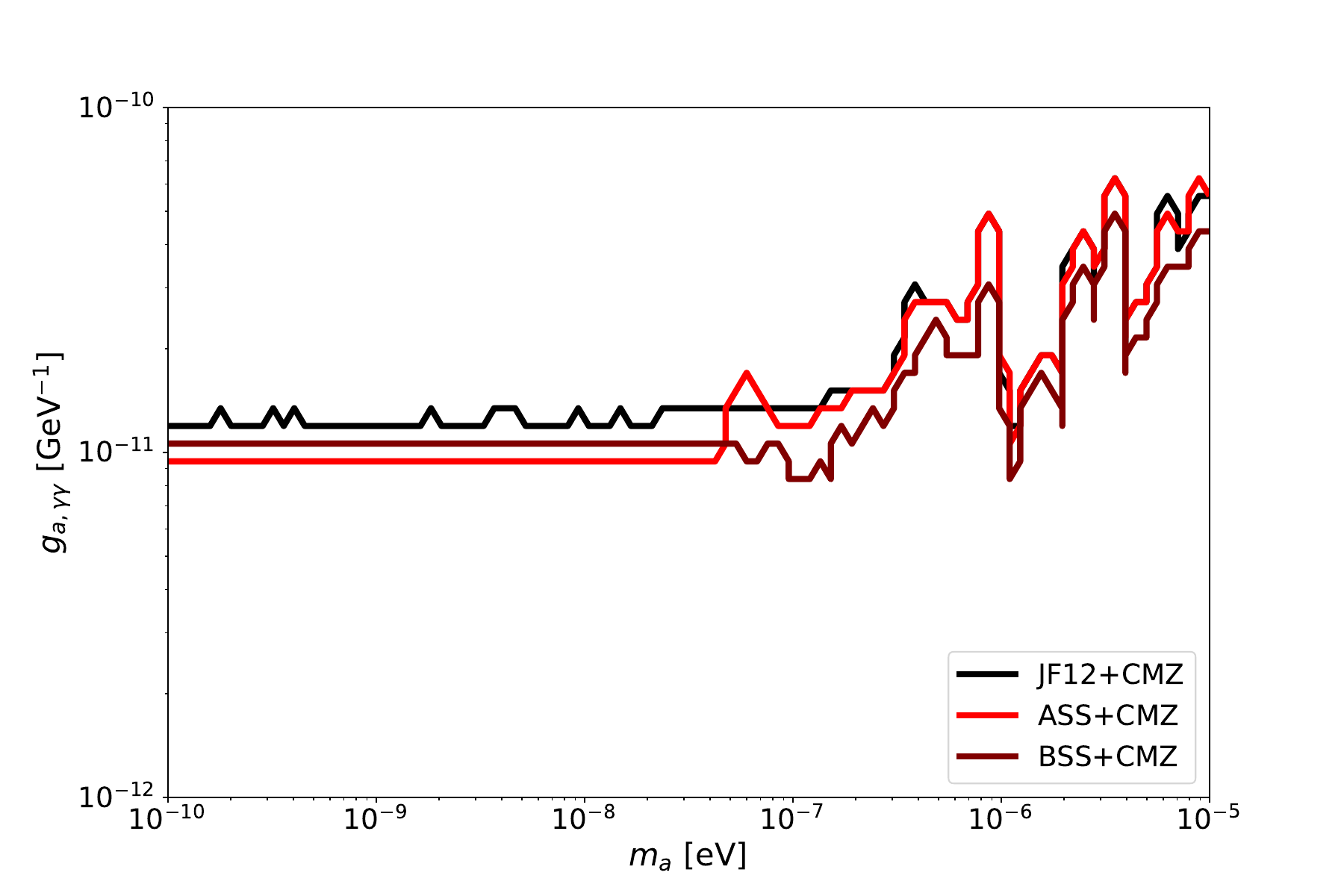}
    \includegraphics[width=0.45\textwidth]{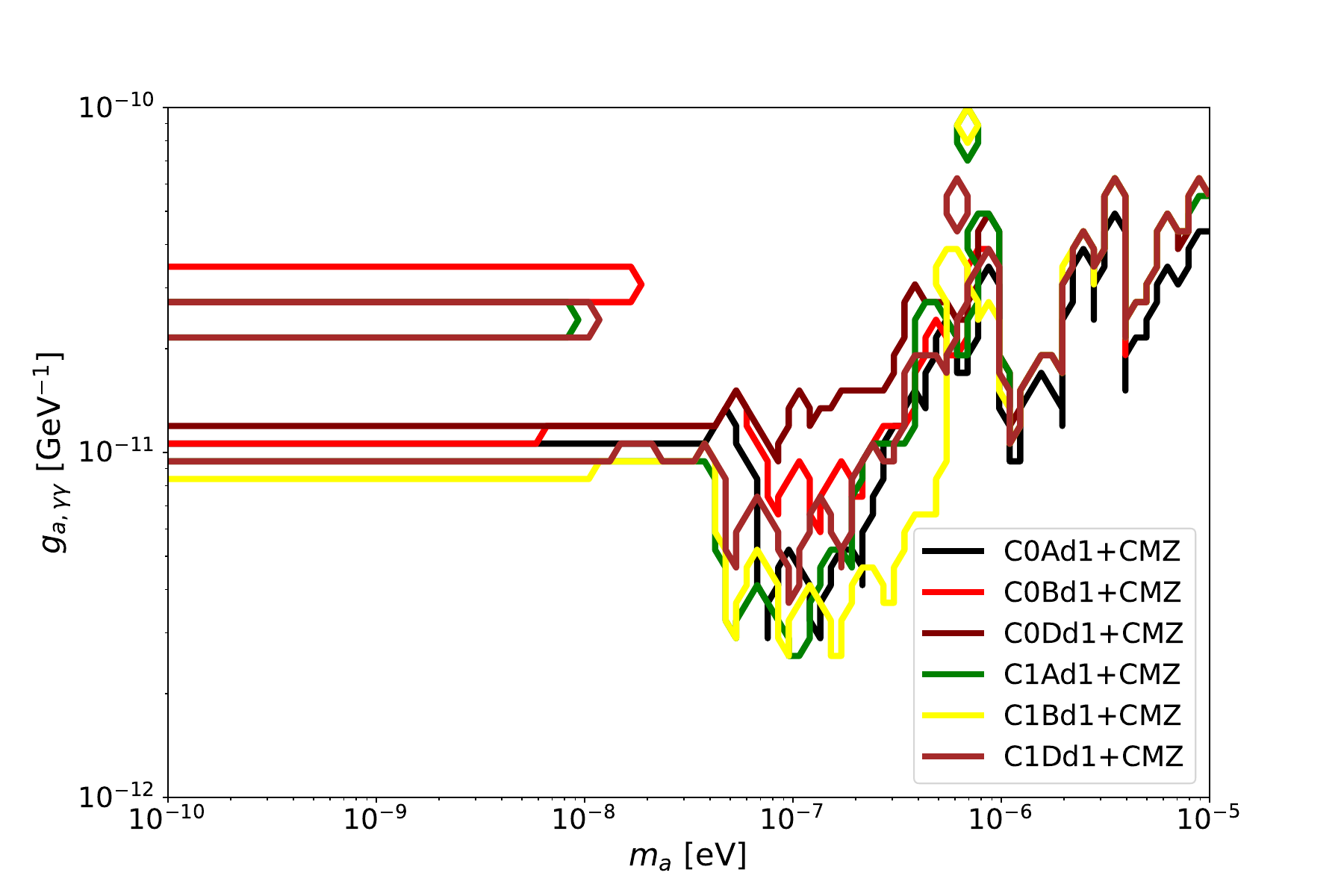}x
    \caption{The constraints derived using various GMF models. Upper panel: results derived from the GMF model (JF12) proposed in Ref.~\cite{J12} and two models (ASS and BSS) proposed in Ref.~\cite{pshirkov}. Lower panel: results derived from six models (C0Ad1, C0Bd1, C0Dd1, C1Ad1, C1Bd1, and C1Dd1) proposed in Ref.~\cite{TF17}.}
 \label{total}
\end{figure}

The constraints on the ALP parameters are influenced by the strength and structure of magnetic fields. Due to the limited understanding of the GMF, the specific GMF configuration may introduce considerable uncertainties. To provide a comprehensive comparison, we establish constraints utilizing eight distinct GMF models obtained from Ref.\cite{pshirkov, TF17} to elucidate the impact of magnetic field models extending beyond the GC region. The comparative analysis is visually 
represented in Fig.\ref{total}. The constraints obtained from various GMF models demonstrate similarities, with the exception of a discrepancy near  $(m_a,g_{a\gamma})\sim (10^{-7}{\ \rm eV},10^{-11}{\ \rm GeV^{-1}})$.
This discrepancy implies a distinct impact of the GMF model in the outer regions on this specific parameter space.  

\begin{figure}
    \centering
    \includegraphics[width=0.45\textwidth]{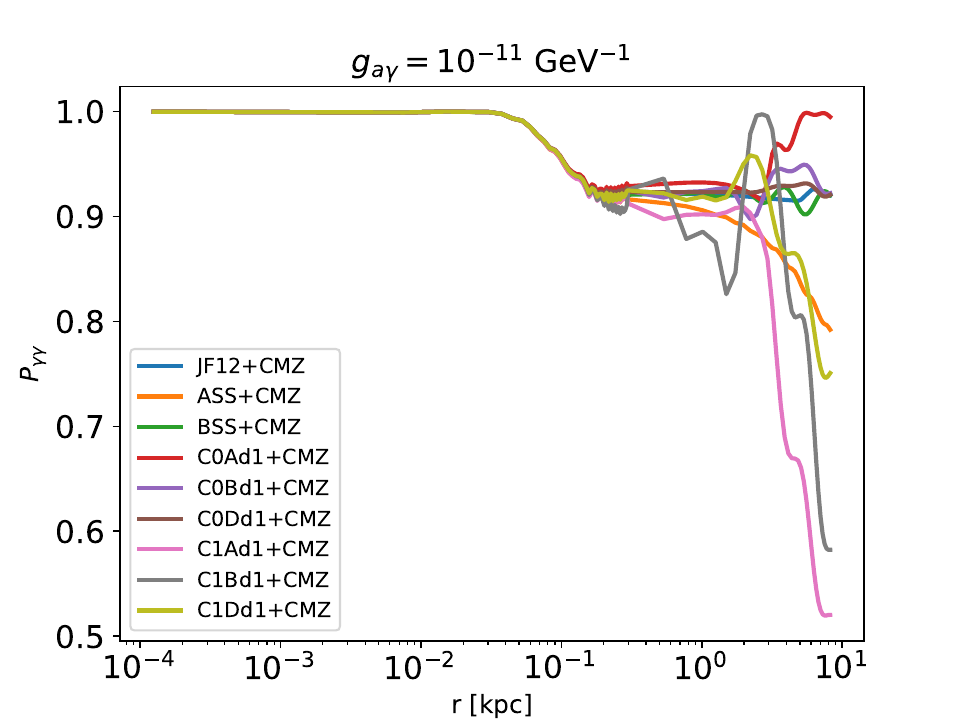}
    \caption{The survival probabilities ${ P_{\gamma\gamma}}$ of photons with an energy of $0.8~$TeV along the propagation direction, for various GMF models at $(m_a,g_{a\gamma})= (10^{-7}{\ \rm eV},10^{-11}{\ \rm GeV^{-1}})$.}
   \label{p_along_los}
\end{figure}

As previously discussed, in the low mass region with $m_a \ll 10^{-7}$ eV, the impact of $\Delta_a$ is suppressed compared to $\Delta_\parallel$, due to the intense magnetic field in the innermost GC region. However, for higher values of $m_a$ and suitable $g_{a\gamma}$, the terms $\Delta_\parallel,\ \Delta_{a\gamma}$ and $\Delta_a$ may collectively affect the mixing matrix. 
In proximity to $(m_a,g_{a\gamma})\sim (10^{-7}{\ \rm eV},10^{-11}{\ \rm GeV^{-1}})$, the impact of the mass term $\Delta_a$ becomes significant, leading to a deviation in the constraints from being solely governed by $\Delta_{a\gamma}$. 
Therefore, in this parameter region, the constraints become sensitive to variations in magnetic field configurations, coupling strengths, and mass parameters. The survival probabilities of photons in this distinctive parameter region for two different GMF models are illustrated in the lower left panel of Fig.~\ref{ratio}, with $m_a$ set at $10^{-7}{\rm\ eV}$. Notably, the oscillations derived with the two GMF models exhibit distinguished disparities, accounting for the structural intricacies presented in the constraints. This feature is further supported by Fig.~\ref{p_along_los}, where the survival probabilities ${\rm P_{\gamma\gamma}}$ of photons with an energy of $0.8~$TeV along the propagation direction for various GMF models are displayed 
at $(m_a,g_{a\gamma})= (10^{-7}{\ \rm eV},10^{-11}{\ \rm GeV^{-1}})$.
These results suggest that the GMF models significantly influence ALP-Photon conversion in this parameter region. Conversely, in the mass region with suppressed $\Delta_a$, the oscillatory patterns in the two models demonstrate relative consistency as depicted in the lower right panel of Fig.~\ref{ratio},
thereby yielding similar constraints.

\begin{figure}
    \centering
    \includegraphics[width=0.45\textwidth]{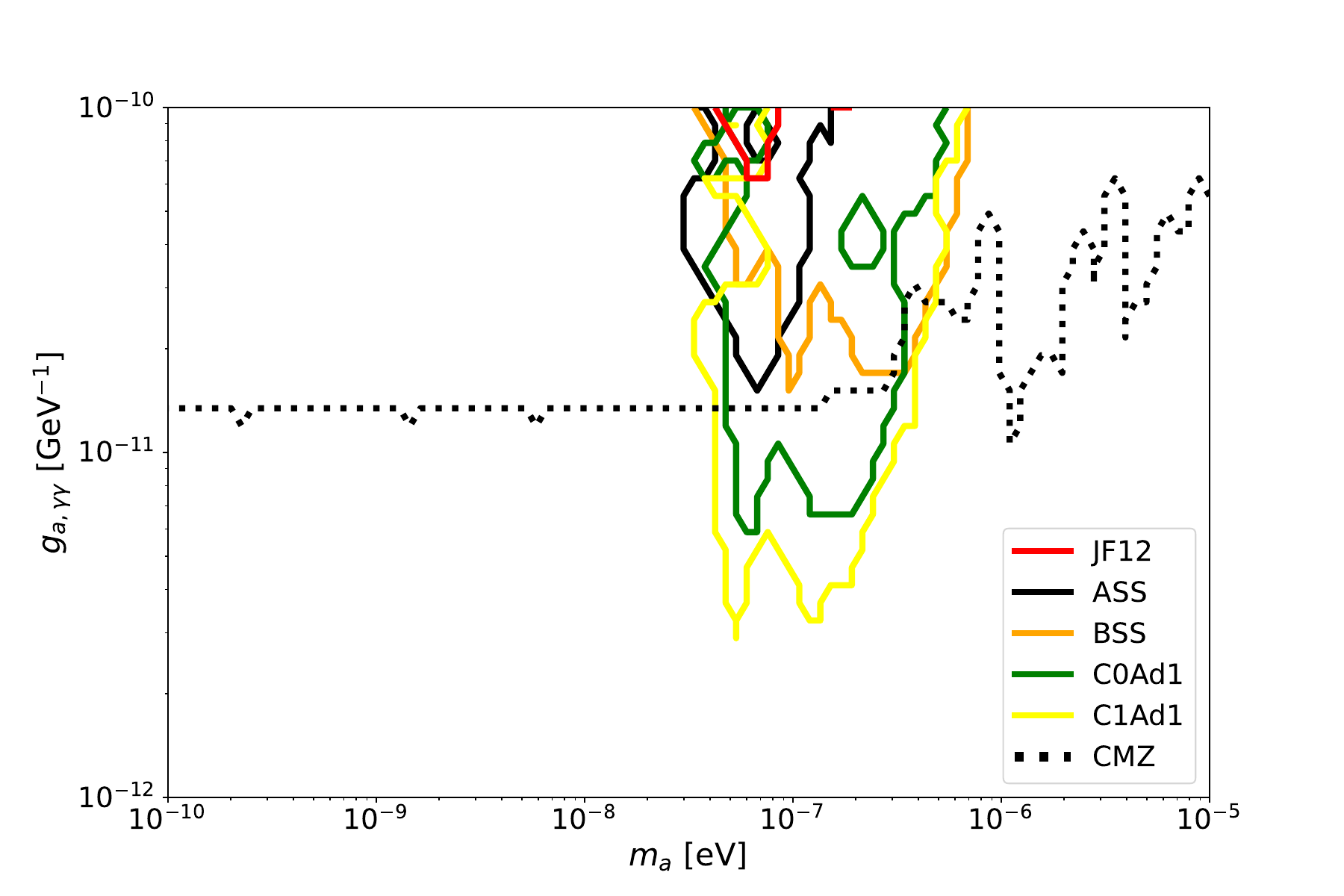}
    \caption{The constraints derived exclusively from the magnetic field in the CMZ and from various GMF models.}
   \label{constraints_exclusively}
\end{figure}

The constraints derived exclusively from the magnetic field in the CMZ and from various GMF models are depicted in Fig.~\ref{constraints_exclusively}. The results derived from the exclusive consideration of the CMZ magnetic field align with the constraints established when incorporating the GMFs in the outer regions. Conversely, the constraints derived from the magnetic field without contributions from the CMZ exhibit notable weakness, showing sensitivity primarily to masses in the vicinity of $\sim10^{-7}{\rm\ eV}$. This implies that the most stringent constraints are determined by the magnetic field present in the GC region. Given that the JF12 GMF model does not encompass the magnetic field within a radius of 3 kpc from the GC in the Galactic disc, it only imposes very weak constraints on the ALP parameters. In contrast, the GMF models in Ref.~\cite{TF17} exhibit a sizable magnetic field strength along the direction of photon propagation, enabling them to establish stringent constraints around $\sim10^{-7}{\rm\ eV}$, which are even more strict than those derived exclusively by the CMZ magnetic field.




\begin{figure}
    \centering
    \includegraphics[width=0.45\textwidth]{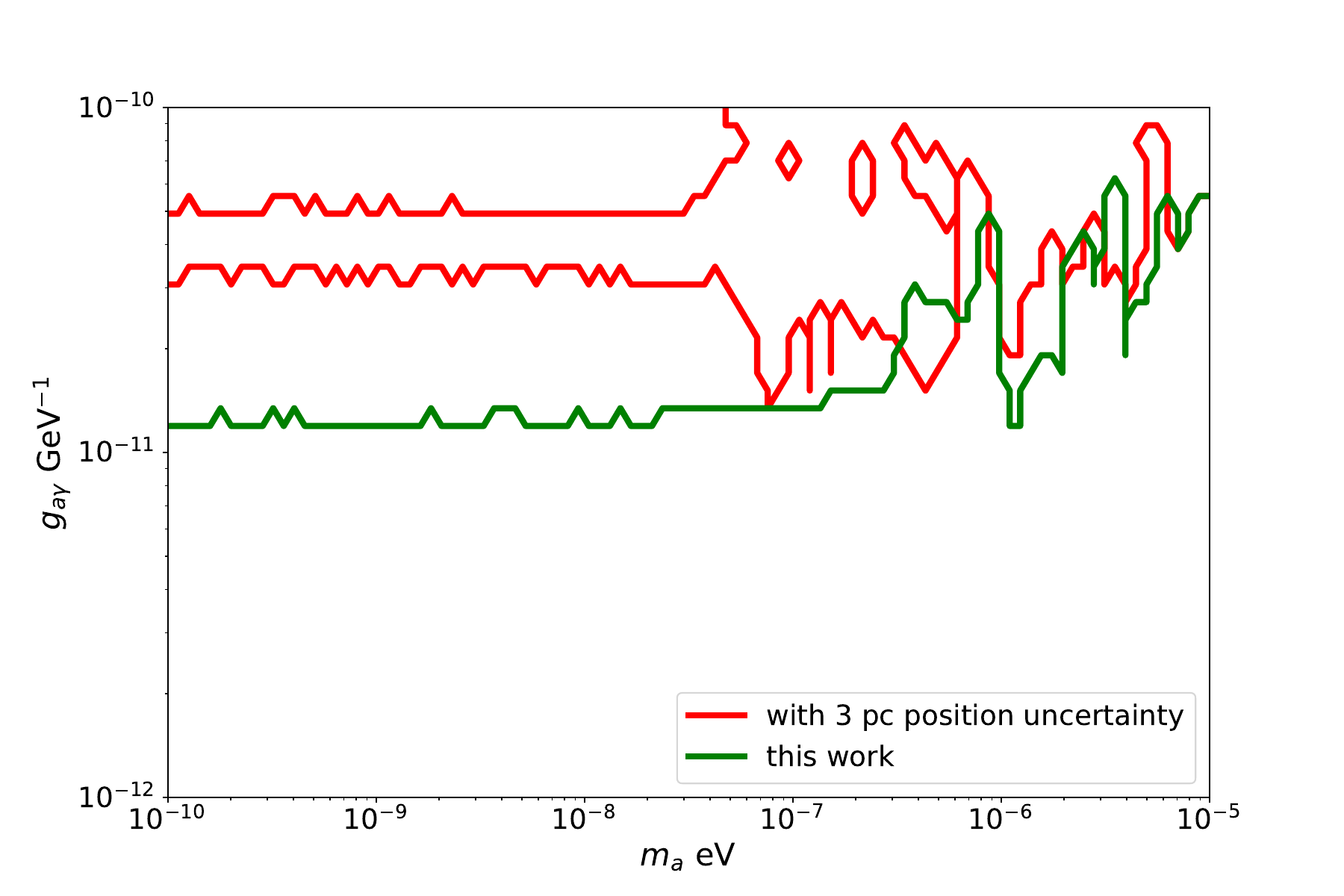}
    \caption{The constraints accounting for the uncertainty related to the location of initial emissions. The constraints are calculated using $P_{\gamma\gamma}$ along the line of sight starting from a distance of 3 pc away from the central point source.}
 \label{pos}
\end{figure}

The gamma-ray spectrum of HESS J1745-290 can be interpreted by leptonic \cite{leptonic}, hadronic \cite{GCdiff}, or hybrid \cite{hybrid} models. However, the exact radiation mechanism and the precise radiation region remain uncertain. 
It is essential to consider the positional uncertainty of the radiation region, as the magnetic field configuration in the CMZ can vary rapidly over small regions. As reported in Ref.~\cite{possgrA}, the upper limit on the spatial extension of HESS J1745-290 is $\sim$3 pc. According to this observation, we consider a positional uncertainty of $\sim$3 pc to assess the potential influence.
For simplicity, we calculate the survival probability $P_{\gamma\gamma}$ along the line of sight starting from a distance of 3 pc away from the central point source. This consideration leads to a reduction in the contribution of the magnetic field in the CMZ. The corresponding constraints are depicted in Fig.~\ref{pos}. Notably, the most stringent constraints in the low-mass region have been weakened as a consequence of this positional uncertainty consideration.

In this study, we adopt the electron density model presented in Ref.~\cite{YMW16} as the baseline model. To evaluate the uncertainties associated with the electron number density $n_{\rm el}$, we also derive constraints utilizing the electron density model proposed in Ref.~\cite{NE2001}. Given the large uncertainty associated with the electron number density in the GC,  we conduct additional tests by manually increasing $n_{\rm el}$, with values even reaching 1000 cm$^3$. The results obtained from these tests exhibit consistency.
Since $\Delta_{\parallel}$ is predominantly influenced by $\Delta_{\rm QED}$, owing to the intense magnetic field in the inner GC region, the electron number density $n_{\rm el}$ exerts a negligible impact on the mixing matrix. As a result, variations in the electron number density do not significantly alter the constraints established in our study.


\section{conclusion} 
\label{sec5}

In this study, we investigate the effect of photon-ALP oscillations in the high energy gamma-ray spectrum 
of HESS J1745-290, which is the point source potentially associated with the supermassive black hole in the GC. The GC is recognized as one of the most complex and active regions in the Milky Way. With a high concentration of molecular clouds and non-thermal filaments within the innermost 200 pc, this region hosts a notably intense magnetic field. The model proposed in Ref.~~\cite{Guenduez2020} has provided valuable insights into the magnetic field configuration in this region, which can significantly influence the phenomenon of photon-ALP oscillations in the VHE gamma-ray spectrum.

By combining this strong magnetic field model in the CMZ with outer GMF models, we have established significant constraints on the ALP parameters. In the regime where the ALP mass is less than approximately $\mathcal{O}(10^{-8}$) eV, the constraints exhibit minimal reliance on the ALP mass. The constraints derived in this study surpass the CAST limits across a broad range of ALP masses. Notably, our results indicate that the constraints derived from HESS J1745-290, a Galactic source, can complement the constraints obtained from gamma-ray observations of extragalactic sources over a wide range of parameters. The uncertainties associated with the GMF  models have a minor impact on these results. This can be primarily attributed to the dominant influence of the magnetic field strength within the innermost GC region, which remains robust despite these uncertainties. Future observations in this region with higher precision hold the promise of broadening the constraints into a more wide parameter space.

 

\begin{acknowledgments}
We thank Zi-Qing Xia for the very helpful discussion. This work is supported by the
National Key Research and Development Program of China (2022YFF0503304), the National Natural Science Foundation of China (Nos. 12322302, 12227804), the Project for Young Scientists in Basic Research 
of Chinese Academy of Sciences (No. YSBR-061), and the Chinese Academy of Sciences.
\end{acknowledgments}

\bibliographystyle{apsrev}

\bibliography{refer.bib}

\widetext

\end{document}